# Optics-free focusing down to micrometer spot size and spectral filtering of XUV harmonics

K. Veyrinas,[1,*] C. Valentin,[1] D. Descamps,[1] C. Péjot,[1] F. Burgy,[1] F. Catoire,[1] E. Constant,[2] and E. Mével[1]

[1]Centre Lasers Intenses et Applications, Université de Bordeaux-CNRS-CEA, 33405 Talence Cedex, France
[2]Institut Lumière Matière, Université Claude Bernard Lyon 1-CNRS, 69622 Villeurbanne Cedex, France
[*]Corresponding author: kevin.veyrinas@u-bordeaux.fr

Controlling the wavefront of an extreme ultraviolet (XUV) high-order harmonic beam during the generation process offers to focus the beam without resorting to any XUV optics. By characterizing the XUV intensity profile and wavefront, we quantitatively retrieve both the size and the position of the waist of each generated harmonics and show that optics-free focusing leads to focused XUV spot with micrometer size. We use this remarkable coherent effect to demonstrate efficient and adjustable spectral filtering of the XUV light, along with a strong rejection of the fundamental beam, without using any XUV optics.

High-order harmonic generation (HHG) [1,2] is a highly non-linear process producing coherent light in the extreme ultraviolet (XUV) domain and is nowadays commonly used as a table-top source of attosecond pulses [3,4]. Numerous applications rely on these HHG-based sources such as ultrafast molecular dynamics [5–7], coherent diffraction imaging [8,9], non-linear optics with XUV photons [10,11], for instance. A major challenge is now to increase the XUV intensity in order to perform XUV-pump-XUV-probe experiments. This can be done by increasing the energy, shortening the duration of the XUV pulse, or by controlling the focusing properties of the XUV beam which strongly depend on the harmonic order [12–14].

For a given laser system, the pulse energy and duration define the typical XUV flux that can be achieved after optimization. The XUV intensity can then be increased by focusing the XUV beam provided the focusing system does not induce significant pulse distortion [15]. Techniques based on reflective mirrors have been used as metallic toroidal mirror [16] or Mo/Si parabolic mirror [17] but they can induce optical aberrations leading to pulse broadening [15]. Multilayer mirrors specifically designed for attosecond pulse-shortening can also be used but suffer from low reflectivity and lack of versatility [18]. Compressors based on XUV gratings have now a good transmission when conical diffraction is used but they require large setup.

Recently, Quintard *et al.* have demonstrated the possibility of generating harmonics as either diverging or converging beams showing the qualitative ability to control the XUV beam properties [19]. This so-called optics-free focusing results from the possibility of controlling the XUV phase front inside the generating medium. For a thin generating medium where phase matching effects lead to a global prefactor [20], the spatial phase $\varphi_q(t,r)$ of a given harmonic $q$ can be approximated by:

$$\varphi_q(t,r) = q\varphi_{IR}(r) - \alpha_q(\tau)I_{IR}(t,r) \qquad (1)$$

The first term corresponds to $q$ times the spatial phase $\varphi_{IR}(r)$ of the driving beam. When only this term is considered, the XUV and IR wavefronts are identical. The second contribution, proportional to the IR intensity $I_{IR}(t,r)$, is an atomic phase intrinsic to the HHG process. The $\alpha_q(\tau)$ coefficient depends on both $q$ and $\tau$, the excursion time of the electronic wavepacket in the continuum [21,22]. The spatial evolution of $\varphi_q(t,r)$ changes the curvature of the XUV wavefront. From an equivalent optical perspective, the two above-mentioned contributions can be expressed in terms of the radii of curvature $R_{IR}$ and $R_{atom}$, respectively, and the total radius of curvature $R$ is given by $1/R = 1/R_{IR} + 1/R_{atom}$. Both terms can have similar values. While $R_{atom}$ is always positive, $R_{IR}$ can either be negative (when the IR beam is converging, *i.e.* before the IR focus) or positive (when the IR beam is diverging, *i.e.* after the IR focus). It is thus possible to control the curvature of the XUV wavefront by controlling the interaction parameters such as the IR intensity, the value of $\alpha_q$ and the gas jet position with respect to the IR focus.

In this Letter, we use a spectrally-resolved wavefront characterization of the XUV radiation to retrieve the waist of each generated harmonic for different positions of the gas jet in the laser focus. This constitutes a direct measurement of coherent focusing and we show that (i) foci as small as 5 µm are achieved by controlling the XUV beam without any XUV optics and (ii) the studied harmonics are not all focused at the same position. This reduces the XUV intensity and makes the attosecond temporal structure space-dependent. We also control harmonic optics-free focusing to achieve efficient and adjustable XUV spectral filtering without using optics.

Experiments have been performed on the terawatt Ti:sapphire Eclipse laser facility at CELIA delivering 800-nm, 100-mJ, 40-fs pulses at a 10-Hz repetition rate. The experimental setup has already been described [19]. Briefly, we used 9mJ of laser energy. The beam is spatially filtered to ensure a near-Gaussian spatial profile ($M_{vert}^2 = 1.02$ and $M_{hor}^2 = 1.09$) with a transmission of 50 %. Pulses are then compressed under vacuum down to 40 fs and propagate under vacuum after the compressor. Optimization of the IR wavefront is performed by reflecting the beam on a deformable mirror (HIPAO, ISP System) coupled to an HASO wavefront sensor (Imagine Optic) for characterization, allowing for Strehl ratio higher than 0.97. IR pulses are then focused by a 2-m focal length spherical mirror (peak intensity of $\sim 2 \times 10^{14}$ W/cm$^2$ at focus) in a pulsed neon jet mounted on a translation stage able to move over 15 cm around the laser focus. Negative (resp. positive) value of $z_{jet}$ means that HHG occurs upstream (resp. downstream) of the IR focus. Generated harmonics are then transmitted through the 500-µm-wide entrance slit of an XUV spectrometer before being dispersed by a grazing incidence Hitachi spherical grating (1200 gr/mm) onto an XUV detector consisting of a 40-mm-diameter dual microchannel plates (MCPs) and a phosphor screen imaged by a CCD camera (2.9 m after the laser focus). A movable pinhole with diameter $\varnothing = 140$ µm can be inserted 37 cm after the IR focal plane to characterize the intensity profile and wavefront of each harmonic. At this position, the pinhole can withstand the IR power without damage.

Characterization of the XUV wavefront has been performed using the SWORD (Spectral Wavefront Optical Reconstruction by Diffraction) technique [23]. It consists of recording the diffraction pattern of the XUV radiation through a slit – a pinhole in our case – while scanning the slit across the XUV beam. For each harmonic, the spatial phase of the sampled wavefront slice is extracted from the relative vertical position of the diffraction pattern's centroid with respect to the pinhole position. The corresponding local XUV intensity is obtained by integrating the signal of the diffraction pattern on the MCP. Figure 1 presents the results of the SWORD measurement for harmonics H31, H39 and H47. Fitting the measured XUV intensity profile (a) and reconstructed spatial phase (b) by Gaussian and quadratic functions, respectively, provides the beam size $w_q$ and the radius of curvature $R_q$ of the wavefront in the pinhole plane. We observe that the beam size and wavefront curvature change with the harmonic order and gas jet position. For $z_{jet}$ = -35 mm, we obtain $w$ = 309,

240, and 174 µm and $R = 363$, 396, and 382 mm for harmonics H31, H39 and H47, respectively.

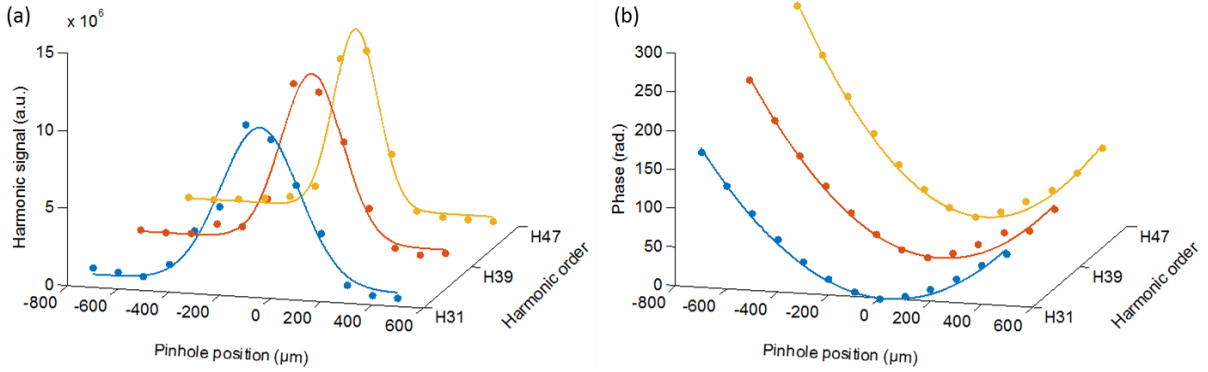

FIG. 1. Measured intensity profile (dots) and Gaussian fit (line) (a) and reconstructed spatial phase of the wavefront (dots) and quadratic fit (line) (b) of harmonics H31 (blue), H39 (orange) and H47 (yellow), using the SWORD technique for $z_{jet} = -35$ mm.

We observe that the spatial profile is close to Gaussian and that the curvature is quadratic. Under these conditions, we can assume that the XUV beam propagates like a Gaussian beam and retrieve, from this intensity profile and wavefront characterization, both the size $w_{q,0}$ and the position $z_{q,0}$ of the waist of the considered harmonic:

$$w_{q,0} = \frac{w_q}{\sqrt{1+\left(\frac{\pi w_q^2}{\lambda_q R_q}\right)^2}} \quad (2)$$

$$z_{q,0} = \frac{R_q}{1+\left(\frac{\lambda_q R_q}{\pi w_q^2}\right)^2} \quad (3)$$

In Fig. 2, we present the results of this analysis for harmonics H31, H33, H49 and H51, as a function of the gas jet position $z_{jet}$ relative to the IR focus. Strong dependences of $w_{q,0}$ (Fig. 2(a)) and $z_{q,0}$ (Fig. 2(b)) with $q$ and $z_{jet}$ are observed and the harmonic waists are separated longitudinally. The waist size varies from ~ 5 µm to ~ 15 µm for H31-H33 and from ~ 5 µm to ~ 20 µm for H49-H51. As the harmonic order increases, the maximum waist size is larger and obtained for generation further upstream of the IR focus. In Fig. 2(b), the gas jet position is denoted by the dashed line and we observe that there is a full range of parameters where the XUV foci are located after the gas jet and are therefore real. XUV waists below (resp. above) this line are located before (resp. after) the generating medium and thus virtual (resp. real). For $z_{jet} > -35$ mm, the waists of harmonics H49 and H51 are virtual and the XUV radiations are emitted as diverging beams. For $z_{jet} < -35$ mm, these waists are real meaning that harmonics H49 and H51 are emitted as converging beams. The waists of harmonics H31 and H33 are real for $z_{jet} < -5$ mm and virtual for $z_{jet} > -5$ mm. We observe that these converging XUV beams are focused to spot size of 5-20 µm at distances up to 120 mm ($4.4 \times z_{R\_IR}$) after the IR focus without using any XUV optics. Overall, these measurements are in good qualitative agreement with the simulated results reported recently by Quintard *et al.* [19] where harmonic optics-free focusing was introduced and observed qualitatively by an inversion of the XUV beam spatial chirp. Our results of $w_{q,0}$ and $z_{q,0}$ for each harmonic constitute a direct and quantitative evidence of coherent focusing in HHG and show that focus size of a few microns can be obtained with real foci. This direct focusing can lead to high XUV intensities. Indeed, even by considering a single harmonic, an 100-nJ energy per harmonic and a 5-fs duration lead to an intensity of $10^{13}$ W/cm$^2$. Higher intensities could be obtained by focusing all harmonics at the same position which should be possible by controlling the IR intensity profile and wavefront.

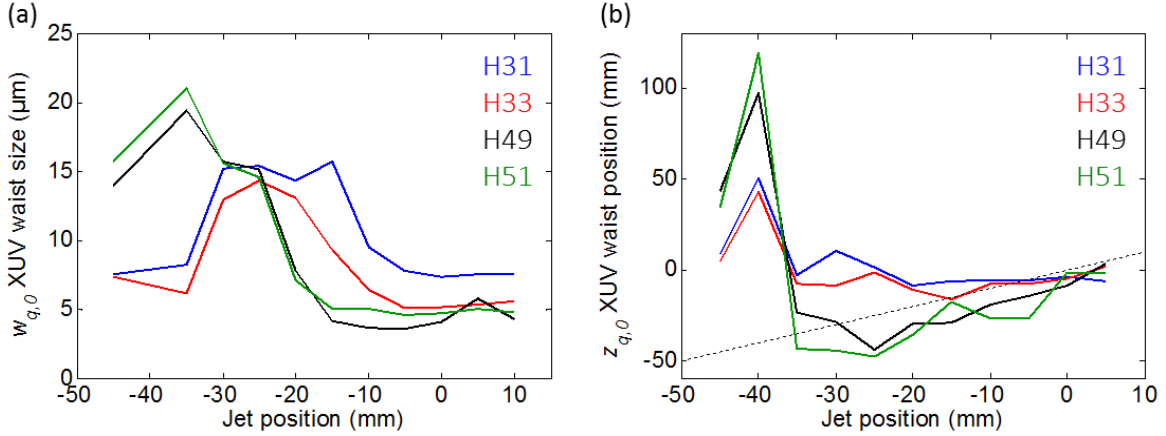

FIG. 2. (a) Size $w_{q,0}$ and (b) position $z_{q,0}$ of the waist of harmonics H31 (blue), H33 (red), H49 (black) and H51 (green) as a function of the gas jet position. In (b), the XUV waists located below (resp. above) the dashed line, that is before (resp. after) the generating medium, are virtual (resp. real).

Spatially-separated foci can also be useful for spectrally-resolved experiments. Here we demonstrate a direct application of the XUV beam properties control by performing XUV spectral filtering without using any XUV optics. This is realized by comparing the spatially-resolved HHG spectrum obtained with a 140-µm-diameter pinhole centered on the pathway of the harmonics and the reference obtained without the pinhole, as shown in Fig. 3(a) and (b) for $z_{jet}$ = -58 mm. The transmission factor (T) is then defined as the ratio of the HHG spectra after integrating over the spatial coordinate (Fig. 3(c)). Figure 3(d) displays the measured transmission factors of harmonics H29 to H47 for $z_{jet}$ = -58, 36, -20 mm and clearly shows that the characteristics of this filter are fully tunable by adjusting the gas jet position with respect to the IR focus. For $z_{jet}$ = -20 mm (black curve), we observe a high transmission (~ 80 %) of harmonics H29-H33 and a good attenuation (T < 20 %) of harmonics H41-H47 (low-pass filter). For $z_{jet}$ = -58 mm (blue curve), the situation is reversed with very high (close to 100 %) and poor (T < 30 %) transmissions of harmonics H45-H47 and H29-H37, respectively. In both cases, the high spectral selectivity and contrast ensure a proper selection of a group of harmonics. On the other hand, for $z_{jet}$ = -36 mm (red curve), broadband transmission is also achieved with 60 < T < 90 % over the H29-H47 range, allowing for producing broadband XUV pulses with similar spatial profiles imposed by the spatial filter. When refocused into an experiment, this can lead to intense attosecond pulses. Finally, this spectral filtering is accompanied with a strong rejection of the IR fundamental beam (T < 1 %). Furthermore, the transmitted IR beam will be strongly diffracted [24] and therefore poorly intercepted by any optics that could be located downstream in an experiment. This can avoid the use of metallic filters that usually reduce significantly the XUV signal. We point out here that the present approach does not require the use of any XUV optics and thus does not suffer from limitations inherent to this energy range and encountered with traditional optical techniques: a pair of gratings offers high spectral selectivity but with limited transmission, while multilayer mirrors have good transmission but are designed for a specific wavelength.

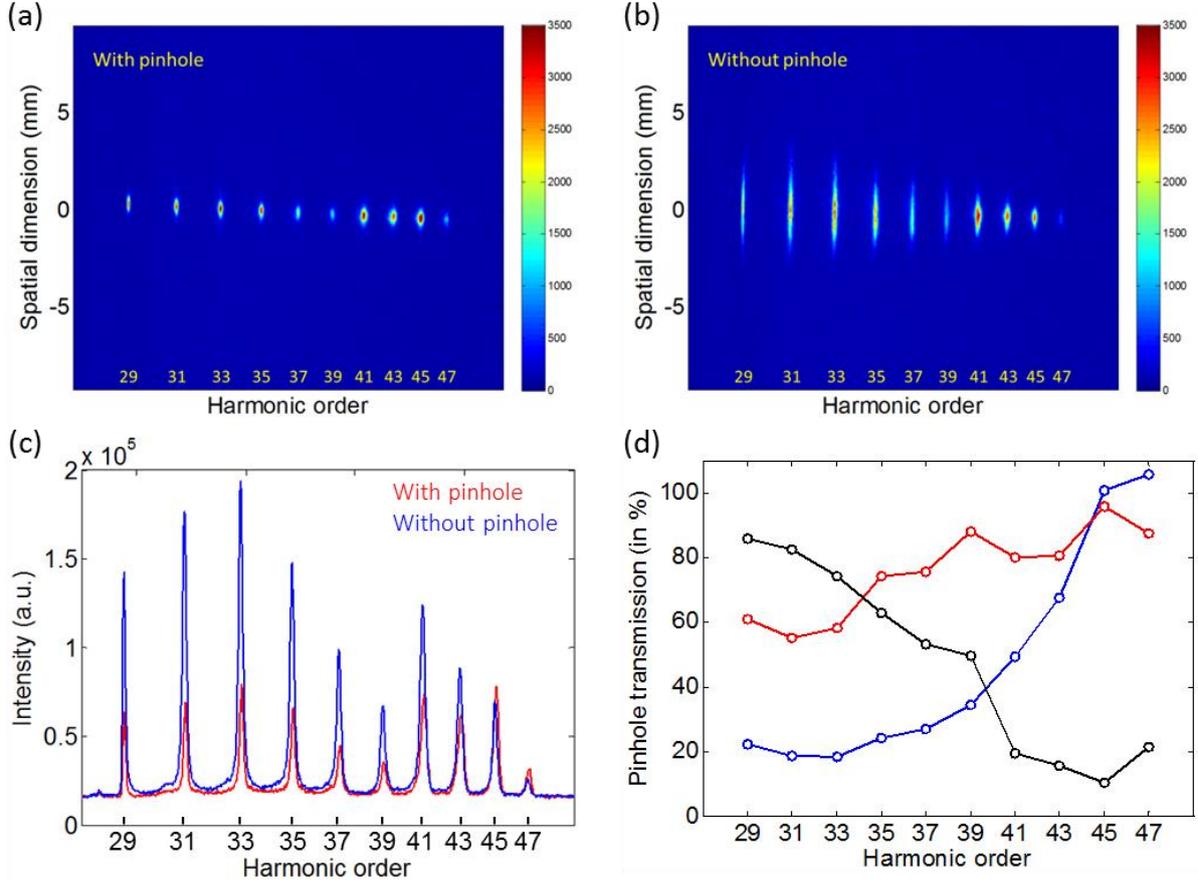

FIG. 3. Far-field spatially-resolved HHG spectrum measured for $z_{jet}$ = -58 mm with (a) and without (b) the pinhole centered on the XUV path. (c) Corresponding HHG spectra after summing over the spatial coordinate. (d) Measured transmission of harmonics H29 to H47 for $z_{jet}$ = -58 (blue), -36 (red) and -20 mm (black).

In conclusion, we investigated the spatial properties of XUV harmonic beams as a function of the position of the generating medium relative to the IR focus, and observed that optics-free focusing of high-order harmonics can be obtained in specific conditions. By means of the SWORD technique, we characterized the XUV intensity profile and spatial phase of the wavefront to retrieve both the size and position of the waists of the generated harmonics. While diverging harmonics are generated with virtual waists as small as 5 μm, converging XUV beams are also produced with spot size of 5-20 μm and can be focused up to 4.4×$z_{R\_IR}$ (120 mm) after the IR focus. We emphasize that no XUV optics are used in this work, thus preserving high photon flux and preventing from optical aberrations that can lead to pulse broadening. This approach is a way to reach higher XUV intensities and should benefit to non-linear XUV-pump-XUV-probe experiments.

Taking advantage of the control of the harmonic focusing properties, we demonstrated adjustable XUV spectral filtering without resorting to any XUV optics. It is characterized by high XUV transmission (60 < T < 100 %), full tunability (low-pass, high-pass or broadband filter) controlled by adjusting the gas jet position with respect to the IR focus, and is accompanied with a strong rejection of the IR fundamental and low-order harmonic beams.

This technique allows narrowband transmission with high spectral selectivity and contrast of particular interest, for instance, in studying molecular dynamics with experiments based on photoelectron spectroscopy, in which we may prefer to reduce the number of harmonics contained in the XUV pulse as the complexity of measured photoelectron spectra increases drastically with the number of ionizing photon energies. On the other hand, spectral filtering also allows broadband XUV transmission and thus supporting the production of focused attosecond pulses.

The next step of this work will consist of shaping the spatial intensity profile of the IR fundamental beam to further control the spatial properties of the generated harmonics. It has been shown that using a flat-top − or super-Gaussian − intensity profile near focus produces very collimated XUV harmonics with a very low divergence associated with a high photon flux [25,26]. In this case, the spatial phase $\varphi_q(t,r)$ of each harmonic given in Eq. (1) would not depend on the harmonic order $q$ leading to a tight focusing over a broad XUV spectral range. This spatial control should directly affect the temporal profile of the XUV beam and produce intense and short attosecond pulses.

We acknowledge the financial support from the Agence Nationale de la Recherche (ANR) through the 'CIRCE' project (ANR-16-CE30-0012) and from the Région Nouvelle-Aquitaine through the 'OFIMAX' project.